\documentclass[12pt]{article}

\newtheorem{proposition}{Proposition}
\newtheorem{lemma}{Lemma}
\usepackage{vmargin}
\setmarginsrb{3.2cm}{2.5cm}{3.2cm}{2.5cm}{1cm}{1cm}{1cm}{1.6cm}
\usepackage{amsmath}
\usepackage{amssymb}
\usepackage[dvips]{graphicx}
\usepackage{rotating}
\usepackage{psfrag}
\usepackage{fancyhdr}

\begin{document}
\begin{center}

{\Large \bf Hamiltonian Analysis of Plebanski Theory
}

\bigskip
\bigskip
\bigskip

{ \sc E. Buffenoir${}^a$\footnote{\tt buffenoi@lpm.univ-montp2.fr},
 M. Henneaux${}^b$\footnote{\tt henneaux@ulb.ac.be }, K.Noui${}^c$\footnote{\tt noui@gravity.psu.edu }, Ph.Roche${}^d$\footnote{\tt roche@lpm.univ-montp2.fr} 
 \\[1cm]}

{\it $^{a,d}$ CNRS, Laboratoire de Physique Math\'ematique et Th\'eorique, 
Universit\'e Montpellier II, 34095 Montpellier, France}\\[3mm]

{\it $^b$  Physique Th\'eorique et Math\'ematique and International
Solvay Institutes,\\
 Universit\'e Libre de Bruxelles, B--1050 Bruxelles, Belgium.}\\[3mm]

{\it $^b$ Centro de Estudios Cient\'{\i}ficos (CECS), Valdivia, Chile.}\\[3mm]

{\it $^c$  Center for Gravitational Physics and Geometry,\\
Penn State University, State College, PA 16801, USA.}\\[3mm]

\vspace*{1.5cm}

\large{\bf Abstract}
\end{center}
We study the Hamiltonian formulation of Plebanski theory in both
the Euclidean and Lorentzian cases. A careful analysis of the
constraints shows that the system is  non regular, i.e. the rank
of the Dirac matrix is non-constant on the non-reduced phase
space. We identify the gravitational and topological sectors which
are regular sub-spaces of the non-reduced phase space. The theory
can be restricted to the regular subspace which contains the
gravitational sector. We explicitly identify first and second
class constraints in this case. We compute the  determinant of the
Dirac matrix and the natural measure for the path integral of the
Plebanski theory (restricted to the gravitational sector). This
measure is the analogue of the Leutwyler-Fradkin-Vilkovisky
measure of quantum gravity.

\noindent

\section*{I. Introduction}
Plebanski theory \cite{Plebanski} is a $4$-dimensional $BF$ theory
with an additional field which forces the $B$ field to satisfy the
simplicity constraint. By Plebanski theory we mean here the real
 formulation described by the real action defined 
below (\ref{Plebanskiaction}) and not the complex formulation
 whose properties have been investigated in \cite{Jacob1,Jacob2}.   It contains, as a particular sector,
$4$-dimensional pure gravity and is therefore an interesting field
theory.  The quantum properties of this field theory are, however,
largely unknown. One important line of study aims at discretizing
this quantum field theory with the tools of lattice gauge field
theory leading to spin-foam models. Although spin-foam models have
been the subject of numerous works over these last years (see the
introduction \cite{Baez} and the review \cite{Perez}), central
issues are not understood and the technical tools needed to
address these central questions need still to be developed. In
particular we had in mind two pressing questions when beginning
this work:

-is it possible to compute from first principle the weight of the
faces, edges and vertices in the spin foam model description  of
Plebanski theory?

-can we see the appearance of quantum groups in Plebanski theory
with cosmological constant?

The following work is a study of the Hamiltonian description of
Plebanski theory. In particular we want to address the following
problems:

-computation of the Liouville measure in the path integral expressed in  term of the original variables of Plebanski theory. This could
be a first step for understanding how to fix the measure of spin
foam models.

-computation of the Dirac bracket of all the  fields once all
second class constraints have been taken into account. In
particular one could have expected to find some brackets of
quadratic type which are the signature of quantum groups.
Unfortunately the Dirac brackets of the variables are complicated
and no hint of quantum groups have appeared yet using this method.

Our paper is organized as follows.  In the next section (section
II), we briefly recall the Plebanski description of gravity as a
constrained $B F$ theory. We then give the Hamiltonian analysis of
the theory (primary and secondary constraints, first and second
class constraints)(section III).  We finally derive the path
integral measure (section IV).

\section*{II. Gravity as a Constrained Topological Field Theory}

This section is devoted to briefly reviewing the classical
Lagrangian analysis of $BF$ and Plebanski theories, and to
presenting the notations that we are using throughout the work.
Let ${\cal M}$ be  a four dimensional oriented smooth manifold
$\cal M$ and $\Omega^1({\cal M})$ the space of one-forms defined
on the manifold $\cal M$. A local coordinate system will be
denoted by $x^{\mu}$, with $\mu\in \{0,1,2,3\}$.

\subsection*{II.1. BF action with cosmological term}
\subsubsection*{II.1.1. Dynamics}
{}For any Lie group $G$ one can define a $BF$ theory which is a
topological quantum field theory \cite{Ho,Birmingham,Cattaneo}. In
the sequel we will only be interested in the case where the group
$G= SO(\eta)$ with $\eta=\text{diag}(\sigma^2,1,1,1)$ and
$\sigma=1$ (resp. $\sigma=i$). We will coin the name Riemannian
(resp. Lorentzian) $BF$-theory in these cases. As usual, we will
denote  $\mathfrak g= so(\eta)$ the Lie algebra of the group whose
basic properties are recalled in the appendix A. By convention,
given two $\mathfrak g$-valued one forms $\alpha=\alpha_\mu
dx^\mu$ and $\beta=\beta_\mu dx^\mu$, their commutator is defined
to be the $\mathfrak g$-valued two-form
$[\alpha,\beta]=[\alpha_\mu,\beta_\nu]dx^\mu \wedge dx^\nu$
whereas the exterior product of these forms is given by $\alpha
\wedge \beta = \frac{1}{2}[\alpha,\beta]$. We will also make an
extensive use of the symmetrization notation:
$\gamma_{(\mu\nu)}=\frac{1}{2}(\gamma_{\mu\nu}+\gamma_{\nu\mu})$
for any matrix $\gamma$.

The action defining $BF$-theory with a cosmological constant
$\Lambda$ is a functional $S_{BF}[A,B]$ of a $\mathfrak g$-valued
two-form $B=\frac{1}{2}B_{\mu \nu} dx^\mu \wedge dx^\nu$, the
$B$-field, and a $\mathfrak g$-connection $A=A_{\mu} dx^{\mu}$
given by:
\begin{eqnarray}\label{BFaction}
S_{BF}[A,B]  =  \frac{1}{2} \int_{\cal M} \!\! d^4x \;
\epsilon^{\mu \nu \rho \sigma} \left(<B_{\mu \nu},F_{\rho \sigma}> +
\frac{\Lambda}{2} \prec B_{\mu \nu}, B_{\rho \sigma}\succ \right)
\end{eqnarray}
where $F= dA + A \wedge A = \frac{1}{2}F_{\mu \nu} dx^\mu \wedge
dx^\nu$ is the curvature of the connection $A$, $\epsilon^{\mu \nu
\rho \sigma}$ is the Levi-Civita symbol and the two independent
Killing forms on the Lie algebra, $<\! \cdot, \cdot \!>$ and
$\prec \! \cdot, \cdot \!\succ$, are precisely defined in the
appendix A (see (\ref{Killingforms})). As explained in the next
section, the choice of the second Killing form in the cosmological
term is made so that the full Plebanski action, obtained by adding
the so-called simplicity constraints to (\ref{BFaction}), is
equivalent (in the appropriate sector) to ordinary gravity.

Although this theory is well-known to be a topological field
theory (see \cite{Baez} for example), it still contains many
unsettled questions such as the evaluation of the expectation
values of observables (see \cite{Cattaneo} as well as more recent
articles on this subject.)

The equations of motion of $BF$ action with respect to the fields
$B$ and $A$ are respectively given by:
\begin{eqnarray}\label{equationsofmotionofBFtheory}
\iota(F) + \Lambda \; B \; = \; 0 \;\;\;\; \text{and}
\;\;\;\; DB \; \equiv dB + [A,B] \; = \; 0 \;.
\end{eqnarray}
In these equations, $\iota$ denotes the Hodge map  defined in the appendix A.
The situation is therefore quite different in the cases $\Lambda=0$ and
 $\Lambda\not=0.$

The $B$-field is locally covariantly constant but in the case
where $\Lambda \neq 0$, is completely determined once the
connection $A$ is given. Furthermore, in that case, the second
equation (\ref{equationsofmotionofBFtheory}) is an immediate
consequence of the first one and the Bianchi identity. Any
solution of the equations of motion is therefore given by a couple
$(A,B)$ with $A$ any connection and $B$ given by
$B=-\Lambda^{-1}\iota(F).$ When the cosmological constant is zero,
any solution of the equations of motion is given by a couple
$(A,B)$ with  $A$  any flat  connection and  $B$ covariantly
constant. Moreover, as the connection $A$ is flat, the $B$-field
can be locally written as $B=D \omega$ for some $\mathfrak
g$-valued one-form $\omega$.

\subsubsection*{II.1.2 Gauge symmetries}
We now discuss the symmetries of the $BF$-action. First of all, we
have the group of gauge transformations ${\cal G}=C^{\infty}({\cal
M},G)$ whose  action on the fields is given by:
\begin{eqnarray}\label{gaugetransformations}
\forall \; g \in {\cal G}, A^{g} = g^{-1} A g +
g^{-1} dg \;\;\; \text{and} \;\;\;  B^{g} = g^{-1} B g \;.
\end{eqnarray}
The invariance of the action (\ref{BFaction}) under the group of
gauge transformations is an immediate consequence of the
invariance of the Killing forms (\ref{Killingforms}). Besides
these familiar gauge transformations, $BF$-theory admits another
set of symmetries ${\cal T}=\Omega^1({\cal M})\otimes \mathfrak g$
whose action on the fields depends explicitly on the cosmological
constant as follows:
\begin{eqnarray}\label{translationaltransformations}
\forall \; \omega \in {\cal T}, \;\;  A^{\omega} = A -
2\Lambda \iota(\omega) \; \text{and} \;  B^{\omega} = B
+ D\omega - 2 \Lambda \iota([\omega,\omega]) \;.
\end{eqnarray}
The proof of the $BF$-action invariance under these transformations
is a straightforward computation and uses, as a central tool,
the property (\ref{swichproperty}) and the Killing forms invariance.

The symmetry group $\cal S$ of the action (\ref{BFaction}) is
obtaining by combining (\ref{gaugetransformations}) and
(\ref{translationaltransformations}), i.e., is the semidirect
product ${\cal S}= {\cal G} \ltimes {\cal T}$ where the action on
the fields is given by $ (A,B)\lhd (g,\omega)\; =\;
((A^{g})^{\omega}, (B^{g})^{\omega})$. The composition law between
two elements $(g_1,\omega_1)$ and $(g_2,\omega_2)$ of $\cal S$ is
given by:
\begin{eqnarray}\label{compositionlaw}
(g_1,\omega_1) \circ (g_2,\omega_2) \; = \; (g_1 g_2, \omega_2 +
g_2^{-1} \omega_1 g_2)\;.
\end{eqnarray}
At this point, it is interesting to note that, in contrast to the
three-dimensional case \cite{Witten}, the symmetry group structure
(\ref{compositionlaw}) is the same for any value of $\Lambda$. The
cosmological constant appears only in the expression of the action
of the normal subgroup $\cal T$ on the fields
(\ref{translationaltransformations}).

The infinitesimal expression of the symmetries easily follows from
(\ref{gaugetransformations}, \ref{translationaltransformations})
and reads :
\begin{eqnarray}\label{infinitesimalsymmetries}
\delta_{(u,\tau)} A  =  Du - 2\Lambda \iota(\tau) \;\;\;\;
\text{and} \;\;\;\; \delta_{(u,\tau)} B  =  B + [u,B] + D\tau \;,
\end{eqnarray}
where $u \in C^{\infty}({\cal M},\mathfrak g)$ and
$\tau \in \Omega^1({\cal M})\otimes \mathfrak g.$

Moreover, being given by the integral of a 4-form, the $BF$-action
is also invariant under the group of  diffeomorphisms of ${\cal
M}$. In the case of Riemannian or Lorentzian BF theory, there is
an interplay between the action of diffeomorphisms and the action
of the symmetry group ${\cal S}.$ To see that this is indeed the
case, let us compute the Lie derivative ${\cal L}_\xi$ of the
fields $A$ and $B$ along a vector field $\xi=\xi^\mu
\partial_\mu$. An easy calculation shows that these Lie
derivatives are related to the infinitesimal symmetries
(\ref{infinitesimalsymmetries}) as follows:
\begin{eqnarray}
({\cal L}_\xi A)_\mu^{{}_{(\varepsilon) a}}  & = &
(\delta_{(u,\tau)}A)_\mu^{{}_{(\varepsilon) a}}  \;
+ \; \xi^\rho \frac{\delta S_{BF}}{\delta B^{{}_{(\varepsilon) a}}_{\rho \mu}} \\
({\cal L}_\xi B)_{\mu \nu}^{{}_{(\varepsilon) a}}
& = &  (\delta_{(u,\tau)} B)_{\mu \nu}^{{}_{(\varepsilon) a}}
\; - \; \epsilon_{\mu \nu \rho \sigma} \xi^{\rho}
\frac{\delta S_{BF}}{\delta A_\sigma^{{}_{(\varepsilon) a}}} \; ,
\end{eqnarray}
where  $u=\xi^\mu A_\mu$ and $\tau = \xi^\rho B_{\rho \mu}
dx^\mu$. As a result, if the fields $A$ and $B$ satisfy the
equations of motion (\ref{equationsofmotionofBFtheory}), their
transformations under diffeomorphisms can be expressed only in
terms of the symmetries (\ref{infinitesimalsymmetries}), which
form a complete set in the sense of \cite{Henneaux1}.

It follows from the form of the gauge transformations that all the
solutions are locally gauge-related on-shell. When $\Lambda
\not=0$, two solutions of the equations of motion are related by a
symmetry $\omega.$ When $\Lambda=0$, two solutions of the equation
of motion are such that the respective connections are flat, so
they are locally pure gauge and therefore are related by a gauge
transformation. The respective $B$ fields are covariantly
constant, therefore locally there exists $\omega$ which relates
the two.

\subsubsection*{II.1.3. Hamiltonian analysis}

Let us end up this brief review on $BF$ theory by recalling basic
facts \cite{Ho,Baez} about the Hamiltonian analysis (see
\cite{Montesinos} for a recent complete treatment).

{}For this  purpose, we assume that spacetime has the structure
${\cal M}={\mathbb R}\times \Sigma$ where the real line $\mathbb
R$ represents time and $\Sigma$ is an oriented smooth 3-manifold
representing space. We will use a chart $x^\mu$ adapted to this
foliation: $x^0$ is a coordinate on ${\mathbb R}$ whereas $x^{i},
i=1,2,3 $ are coordinates on $\Sigma.$ As usual we denote
$\epsilon^{ijk}=\epsilon^{0ijk}$.

The action (\ref{BFaction}) shows that the spatial components of
the connection $A^{{}_{(\varepsilon) a}}_i$ and $ \epsilon^{ijk}
B_{jk}^{{}^{(\varepsilon) b}} \eta_{ab}= \Pi_{{}^{(\varepsilon)
a}}^i$ are canonically conjugate fields, i.e.
\begin{eqnarray}\label{definitionofPi}
\{A^{{}_{(\varepsilon) a}}_i(x) \; , \;
\Pi_{{}^{(\varepsilon')b}}^j(y) \} \; = \; \delta_a^b \;
\delta_{\varepsilon, \varepsilon'} \;  \delta_i^j \; \delta(x,y)
\;.
\end{eqnarray}
Moreover, a straightforward calculation shows that the
canonical Hamiltonian $H_{BF}[A,B]$ reads:
\begin{eqnarray}\label{BFhamiltonian}
H_{\!BF}[A,B] \; =- \; \int_{\Sigma}\!\! d^3x \;
( \epsilon^{ijk} <B_{0i},F_{jk}> + <A_0,D_i \Pi^i> + \;
\Lambda \prec B_{0i},\Pi^i \succ )\;.
\end{eqnarray}
The components $A^{{}_{(\varepsilon) a}}_0$ of the
connection and the components $B_{0i}^{{}^{(\varepsilon) a}}$
of  the $B$-field  appear to be non-dynamical variables:
they are Lagrange multipliers which impose the following
primary constraints:
\begin{eqnarray}
&& \epsilon^{ijk}\iota(F_{jk}) + \Lambda \Pi^i  =  0\\
&& D_i\Pi^i=0.
\end{eqnarray}
These constraints are obviously contained in the Lagrangian
equations (\ref{equationsofmotionofBFtheory}), are first class and
generate the infinitesimal symmetries. Transformation laws of the
dynamical variables $(A^{{}_{(\varepsilon) a}}_i,
\Pi_{{}^{(\varepsilon) a}}^i)$ under these symmetries are given by
(\ref{infinitesimalsymmetries}) with parameters $u$ and $\tau$
such that $\tau_0=0.$

\subsection*{II.2. Plebanski action}
Plebanski action \cite{Plebanski} is a constrained Riemannian or
Lorentzian BF theory, in which the simplicity constraint on the
$B$ field written explicitly in (\ref{simplicityconditions1})
below is dynamically enforced through the introduction of Lagrange
multipliers.

The Plebanski action is a functional $S_{Pl}[A,B,\varphi]$ of the
$B$-field, the connection  $A$ and Lagrange multipliers $\varphi$.
Actually, there exists two different but
``equivalent''\footnote{To be more precise, the two formulations
are shown to be equivalent on shell when the non degeneracy
condition is satisfied.} formulations for this action
\cite{Freidel} depending on the choice of the Lagrange multipliers
type: $\varphi$ can be a density  tensor field with space-time
indices only  or a field  with Lie-algebra  indices (internal)
only. The latter formulation is more customary in the literature
on spin foams  but we will  use the former which turns out to be
more convenient for the achievement of an  Hamiltonian analysis.
The $SO(\eta)$ Plebanski action is  given by:
\begin{eqnarray}\label{Plebanskiaction}
S^{Pl}[A,B,\varphi] \; = \; S_{BF}[A,B] +
\frac{1}{4} \int_{\cal M} \!\! d^4x \; \varphi^{\mu \nu \rho \sigma}
\prec B_{\mu\nu},B_{\rho \sigma}\succ \;.
\end{eqnarray}
In this expression, $\varphi^{\mu \nu \rho \sigma}=\varphi^{[\mu
\nu][\rho \sigma]}$ is a density tensor, symmetric under the
exchange of the pair $[\mu \nu]$ with $[\rho \sigma]$  and
satisfying the tracelessness condition:
$\epsilon_{\mu\nu\rho\sigma} \varphi^{\mu \nu \rho \sigma}=0$.
Therefore it contains $20$ independent components. Varying the
action (\ref{Plebanskiaction}) with respect to $\varphi$ imposes
the following $20$ algebraic equations on the $36$ components of
the $B$-field:
\begin{eqnarray}\label{simplicityconditions1}
\prec B_{\mu \nu}, B_{\rho \sigma} \succ \; = \; \frac{\cal V}{4!}
\; \epsilon_{\mu \nu \rho \sigma}
\end{eqnarray}
where $\cal V= \epsilon^{\mu \nu \rho \sigma} \prec B_{\mu \nu},
B_{\rho \sigma} \succ.$ In the sequel, $\cal V$ will be called the
four-dimensional volume. When $\cal V$ is not vanishing it is said
that the non degeneracy condition is satisfied. The equations
(\ref{simplicityconditions1}) are known  as ``simplicity
constraints'' and their properties and solutions have been studied
and classified when the non degeneracy condition is satisfied
\cite{Reisenberger,Depietri,Freidel}. In particular, it was shown
that they are equivalent to the following system of equations:
\begin{eqnarray}\label{simplicityconditions2}
\epsilon^{\mu \nu \rho \sigma} B_{\mu \nu}^{IJ} B_{\rho
\sigma}^{KL} \; = \; \frac{\cal V}{4!} \; \epsilon^{IJKL}\;.
\end{eqnarray}
Moreover, simplicity conditions and the non degeneracy conditions
are simultaneously fullfilled  if and only if there exists a non
degenerate cotetrad field $e^I=e^I_\mu dx^\mu$ such that the
$B$-field can be written in one of the following forms:
\begin{eqnarray}
 \text{ {\it topological sector} (I$\pm$})& : & B^{IJ} \;
 = \; \pm \; e^I \wedge e^J \\
\text{ {\it gravitational sector} (II$\pm$)} & : & B^{IJ}
\; = \; \pm \; \frac{1}{2} \epsilon^{IJ}{}_{KL} \;
e^K \wedge e^L \;.\label{solutionsofsimplicityconstraints}
\end{eqnarray}
We now come to the relation between Plebanski theory and Palatini
formulation of general relativity. If one substitutes the
solutions of the sector $(II\pm)$
(\ref{solutionsofsimplicityconstraints}) into the Plebanski
action, one obtains  $\pm S^{Pa}_{ \pm\Lambda}[A,e]$, where $
S^{Pa}_{ \Lambda}[A,e]$ is the Palatini action of general
relativity with cosmological constant $\Lambda:$
\begin{eqnarray}
S^{Pa}_{\Lambda}[A,e] \; = \; \frac{1}{4}\int_{\cal M}
\!\! d^4x \; \epsilon^{\mu \nu \rho \sigma}  \;
\epsilon_{IJKL} \; e^I_\mu \; e^J_\nu \; \left( F^{KL}_{\rho \sigma}
+ \frac{\Lambda}{4} \; e^K_\rho \; e^L_\sigma \right) \;.
\end{eqnarray}
This result gives an a posteriori justification of the choice for
the Killing form defining the cosmological term of the $BF$-action
(\ref{BFaction}).  The topological sector however  does not have
any physical interpretation.

Note that when the non degeneracy condition is fulfilled the
different set of solutions of the simplicity conditions are
disjoint. To see that this is the case, the Urbantke metrics
$g_{\mu \nu}^{(\varepsilon)}$, $\varepsilon \in \{+1,-1\}$,
defined as \cite{Urbantke}:
\begin{eqnarray}\label{Urbantkemetric}
g_{\mu \nu}^{(\varepsilon)} \; = \; -\frac{2}{3 \cal V} \;
\epsilon_{abc}\;\epsilon^{\alpha \beta \gamma \delta} \;
B_{\mu \alpha}^{{}_{(\varepsilon)a}} \;
B_{\beta \gamma}^{{}_{(\varepsilon)b}} \;
B_{\delta \nu}^{{}_{(\varepsilon)c}} \;.
\end{eqnarray}
are a useful tool. If the $B$-field satisfies the simplicity
conditions, their expressions simplify drastically and are given
in term of the cotetrad $e_\mu^I$ by the formulae:
\begin{eqnarray}
 \text{ {\it topological sector $(I\pm)$}} & : &
 g_{\mu \nu}^{(\varepsilon)} \; = \; \pm \;
 \varepsilon \; \eta_{IJ} \; e_\mu^I \; e_\nu^J \\
 \text{ {\it gravitational sector $(II\pm)$}} & :
 & g_{\mu \nu}^{(\varepsilon)} \; = \; \pm \; \eta_{IJ} \; e_\mu^I \; e_\nu^J \;.
\end{eqnarray}
As a result, a non-degenerate $B$-field solution of the simplicity
constraints cannot belong to both topological and gravitational
sectors. Moreover, given a $B$-field in the gravitational or
topological sector, the signature of its associated Urbantke
metrics completely fixes its belonging to the $(\pm)$ sectors. The
four sectors  are therefore  disjoint.

\medskip

\section*{III. Hamiltonian analysis of Plebanski theory}
We now study   $SO(\eta)$ Plebanski theory in the Hamiltonian
framework. For this purpose, we will assume that  ${\cal
M}=\mathbb R \times \Sigma$ where $\Sigma$ is a smooth, oriented
three-manifold (without boundary) which will play the role of
space.

The first step consists in deriving the Hamiltonian and the
constraints.  One must then split the constraints into first class
constraints and second class constraints.  A notable feature of
the Plebanski action is that it does not define a regular
Hamiltonian system in the sense that the rank of the Dirac matrix
depends on which type of sectors the $B$-field belongs to.

We provide a set of holonomic constraints which replace the
simplicity constraints and eliminate the topological sector. When
the $B$-field is  non holonomically enforced  to belong to the
gravitational sector $II+$ we can compute the Dirac matrix
determinant in closed form. This will be used in section IV to
formulate a path integral quantization of Plebanski theory in this
sector.

\subsection*{III.1. Constraint analysis}
The Plebanski action (\ref{Plebanskiaction}) can be written, after
an integration by parts, as:
\begin{eqnarray}
&&S_{Pl}= \int dt \int_{\Sigma} \!\! d^3x
(\epsilon^{mnp}<B_{mn},\partial_0 A_p >+
\epsilon^{mnp}<B_{0p},F_{mn}>+ \nonumber\\
&&<A_{0}, D_p B_{mn}>\epsilon^{mnp}+\Lambda \prec
B_{0p}, B_{mn} \succ\epsilon^{mnp}+ \nonumber\\
&& \varphi^{0m0n}
 \prec B_{0m},  B_{0n}\succ +
 \varphi^{0mnp}\prec B_{0m},  B_{np}\succ +\frac{1}{4}
\varphi^{mnrs} \prec B_{mn},  B_{rs}\succ).
\end{eqnarray}
{}From this expression we see that the fields $A_0, B_{0m},
\varphi$ are non dynamical since their time derivatives do not
occur in the Lagrangian. However the fact that the $B_{0m}$'s
appear quadratically prevents us from treating them as Lagrange
multipliers. To handle this problem, we decided to add to the
Lagrangian the term $$<P^i,\partial_0 B_{0i}> - <\mu_i, P^i>
$$ where $P^i$ and $\mu_i$ are new independent variables.
This does not change the dynamics since the new fields are
auxiliary fields, the elimination of which yields back the
original action (by varying with respect to $\mu_i$ and $P^i$, one
gets $P^i = 0$ and $\partial_0 B_{0i}  -  \mu_i = 0$). This
modified action with the new fields is our starting point.

\subsubsection*{III.1.1. Conjugate momenta and primary constraints}
With the new variables, the Plebanski action is in first order
form:
\begin{eqnarray}\label{firstorderPlebanskiaction}
S^{(1)}_{Pl}[A_i,\Pi_i,P_i, B_{0i},A_{0}, \mu_i, \varphi]  =
 \int dt \int_{\Sigma} \!\! d^3x \;(
 <\Pi^i,\partial_0 A_i> + <P^i,\partial_0 B_{0i}>
  - H)
\end{eqnarray}
where we have made the change of variables \begin{equation} \Pi^i
= \epsilon^{mni}B_{mn}\end{equation} and where the Hamiltonian $H$
reads:
\begin{eqnarray}
&&-H=\epsilon^{mnp}<B_{0p},F_{mn}>+<A_0, D_p \Pi^{p}>+
\Lambda \prec B_{0p},\Pi^{p}\succ+\nonumber\\
&&+\varphi^{0m0n}\prec B_{0m}, B_{0n}\succ+
\frac{1}{2} \varphi^{0mnp}\prec B_{0m}, \Pi^{r}\succ\epsilon_{npr}+\nonumber\\
&&+\frac{1}{16}\varphi^{mnrs}\epsilon_{mni}\epsilon_{rsj}\prec \Pi^{i}, \Pi^{j}\succ -<\mu_i, P^i>.
\end{eqnarray}
We will denote ${\cal V}=\prec B_{0j},\Pi^j\succ$ and assume that
the non degeneracy condition holds i.e ${\cal V}\not=0.$ We
conclude immediately that the only dynamical fields are the
spatial components of the connection $A_i,$  their conjugate
momenta $\Pi^i$ as well as the $B_{0i}$ and their momenta $P^i$.
The other fields $A_0$, $\mu_i$ and $\varphi$ do not appear with
time derivatives. Moreover, the action
(\ref{firstorderPlebanskiaction}) is linear in them.  They can,
therefore, be considered as Lagrange multipliers. Thus, we do not
associate conjugate momenta to these variables in order not to
complicate unecessarily the canonical analysis and we start with
the phase space of the $A_i$'s, the $\Pi^i$'s, the $B_{0i}$'s and
the $P^i$'s with following non-zero Poisson brackets :
\begin{eqnarray}
\{A_i^{{}_{(\varepsilon) a}} (x) \; , \; \Pi^j_{{}^{(\varepsilon') b}}(y)\}
& = & \delta^a_b \; \delta_{\varepsilon,\varepsilon'} \; \delta^i_j \; \delta(x,y) \; , \\
\{B_{0i}^{{}_{(\varepsilon) a}} (x) \; , \; P^i_{{}^{(\varepsilon') b}}(y)\}
& = & \delta^a_b \; \delta_{\varepsilon,\varepsilon'} \; \delta^i_j \; \delta(x,y) \;.
\end{eqnarray}

The ``primary constraints" are given by
\begin{eqnarray}
&&P^i_{(\varepsilon)a}\approx 0\label{Pconstraints}\\
&&\Gamma\equiv D_p \Pi^p\approx 0\label{DPiconstraint}\\
&&\Phi(B,B)_{ij}  \equiv \prec B_{0i}, B_{0j} \succ \;
\approx \; 0 \; , \label{BBconstraints}\\
&&\Phi(B,\Pi)^i_j  \equiv \prec B_{0j},\Pi^i \succ - \frac{1}{3}\;
\delta^i_j \; {\cal V}  \; \approx \; 0 \;, \label{BPconstraints}\\
&&\Phi(\Pi,\Pi)^{ij}  \equiv  \prec \Pi^i , \Pi^j \succ  \;
 \approx \; 0 \;\label{PPconstraints}.
\end{eqnarray}
The constraint (\ref{Pconstraints}) is obtained by varying with
respect to $\mu_i$, constraint (\ref{DPiconstraint}) is obtained
by varying with respect to $A_0$ and the simplicity constraints
(\ref{BBconstraints},\ref{BPconstraints},\ref{PPconstraints}) are
obtained by varying with respect to $\varphi.$ The time evolution
of any phase space function $f$ is given by $\dot{f}=\{f,H\}$.

\subsubsection*{III.1.1. Secondary constraints}

We now study the secondary constraints arising from the
conservation of the primary ones.

Conservation of the constraints (\ref{Pconstraints}) under time
evolution implies the following conditions:
\begin{eqnarray}\label{evolutionofP}
\dot{P}^i \; = \; -\epsilon^{ijk}F_{jk} + \sigma \Lambda \;
\iota(\Pi^i) + 2 \sigma \varphi^{0i0j} \; \iota(B_{0j}) + \sigma
\varphi^{0ijk} \epsilon_{jkl} \; \iota(\Pi^l) \; \approx \; 0 \;.
\end{eqnarray}
These 18 equations involve the 6 components $\varphi^{0i0j}$ and
the 8 components $\varphi^{0ijk}$. Due to the invertibility of
${\cal V}$ and the expressions of the primary simplicity
constraints, 14 of these equations completely fix the value of
$\varphi^{0i0j}, \varphi^{0ijk}$ whereas the 4 remaining ones are
secondary constraints given by:
\begin{eqnarray}
C_0 & \equiv & <\dot{P}^i,B_{0i}> \; = \; \epsilon^{ijk}<B_{0i},F_{jk}>
+ \Lambda \prec \Pi^i , B_{0i} \succ  \; \approx \; 0\;,
\label{scalarconstraint}\\
C_i & \equiv &\epsilon_{ijk} <\dot{P}^j, \Pi^k>  \; =  \;
<F_{ij},\Pi^j> \;\approx \; 0\;.\label{vectorialconstraint}
\end{eqnarray}

Conservation of the constraint (\ref{DPiconstraint}) does not
generate any new secondary constraint. To see this it is useful
(and permissible) to redefine the constraint $\Gamma$ as follows:
\begin{equation}
\Upsilon=\Gamma - [P^i,B_{0i}]\approx 0 \;.
\end{equation}
The constraint $\Upsilon$ generates infinitesimal gauge
transformations of type ${\cal G}$ on all the dynamical variables.
As a result, the constraints $\Upsilon$ are conserved under time
evolution because $H$ is invariant under these transformations.

Before studying  the conservation of the simplicity constraints we
prove some relations implied by the  non degeneracy condition. We
denote by $K, L$ the $6\times 6$ matrices  defined by
\begin{eqnarray}
K^{(\varepsilon)a}_{i}=B^{(\varepsilon)a}_{0i},
 K^{(\varepsilon)a}_{3+i}=\Pi_{(\varepsilon)a}^{i},
L_{(\varepsilon)a}^{3+i}=\frac{3\epsilon}
{\sigma\cal V}B^{(\varepsilon)a}_{0i},
 L_{(\varepsilon)a}^{i}=\frac{3\epsilon}
 {\sigma\cal V}\Pi_{(\varepsilon)a}^{i} \;i=1,...,3.
\end{eqnarray}
The non degeneracy condition and the simplicity constraints imply
that $L$ is the  weak left inverse of $K.$ This implies that $L$
is also the right weak inverse of $K$ which reads:
\begin{equation}
\sum_{k}\Pi^{k}_{(\varepsilon)a}B^{(\varepsilon')b}_{0k}+
\Pi^{k}_{(\varepsilon')b}B^{(\varepsilon)a}_{0k}
\approx\frac{\varepsilon}{3\sigma}{\cal V}\delta^{b}_{a}
\delta_{\varepsilon,\varepsilon'}.
\end{equation}

{}First, we consider the set of constraints
(\ref{BBconstraints},\ref{BPconstraints}) whose time derivative is
given by:
\begin{eqnarray}\label{derivativeBBconstraints}
&&\dot{\Phi}(B,B)_{ij} \; = \; 2 \prec \mu_{(i}, B_{0j)} \succ \; \approx \; 0,
\label{dotPhiBB}\\
&&\dot{\Phi}(B,\Pi)^i_j=\prec \mu_{i}, \Pi^j \succ +
2\epsilon^{imn} \prec B_{0j}, D_m B_{0n}\succ +
<A_0, [B_{0j}, \Pi^i]>\nonumber \\
& - & \frac{1}{3}\delta^i_j (\prec \mu_{k}, \Pi^k \succ +
2\epsilon^{kmn} \prec B_{0k}, D_m B_{0n}\succ +
<A_0, [B_{0k}, \Pi^k]>) \approx 0\label{dotPhiBPi}.
\end{eqnarray}
These 14 equations fix  $14$ independant linear combinations among
the $18$ multipliers $\mu_{i}^{(\varepsilon)a}$ and imply no
further constraint on the dynamical variables.  Indeed, let us
introduce the variables $\nu_{i}^{m}$ with
$\mu_{i}^{(\varepsilon)a}=\sum_{m=1}^6\nu_{i}^{m}K^{(\varepsilon)a}_{m}.$
The set of constraints (\ref{dotPhiBB},\ref{dotPhiBPi}) fix  the
combinations $\nu^{3+i}_j+\nu^{3+j}_i $ and $\nu^{i}_j-\frac{1}{3}
\delta^i_j\sum_{k=1}^3\nu^{k}_k$ where $i,j\in\{1,2,3\}.$ {}From
this analysis, we can reorganize the 18 constraints $P^i\approx 0$
in two subsets of constraints:
\begin{eqnarray}
&&\kappa_0=<P^i,B_{0i}>\approx 0\\
&&\kappa_i=\epsilon_{ijk}<P^j,\Pi^k> \approx 0\; \\
&&\Phi(B,P)^i_j=<P^{i},B_{0j}> - \frac{1}{3} \delta^i_j <P^{k},B_{0k}>\approx 0 \\
&& \Phi(P,\Pi)^{ij}=<P^{(i},\Pi^{j)}>\approx 0 .
\end{eqnarray}
We will show later that $\kappa_0,\kappa_i$ are the 4 first class
constraints corresponding to the previous four Lagrange
multipliers among the $\mu$'s that remain arbitrary.

The study of the conservation of the remaining simplicity
constraints (\ref{PPconstraints}) introduces, by contrast, new
constraints in the theory. Indeed, vanishing of the time
derivatives of these constraints imposes the following conditions:
\begin{equation}\label{Tconstraints}
-\frac{1}{4} \dot{\Phi}(\Pi,\Pi){}^{ij} \; \equiv \;
T^{ij} \; = \;  \epsilon^{mn(i} \prec D_m \Pi^{j)},  B_{0n} \succ \; \approx 0 \;.
\end{equation}
This closes the analysis of the secondary constraints.

\subsubsection*{III.1.3. Conservation of secondary constraints}

In order to complete the Dirac algorithm we have to study the
conservation of these secondary constraints. A direct computation
shows that time derivative of the constraints (\ref{Tconstraints})
can be expressed as:
\begin{eqnarray}\label{conservationoftorsion}
&&\dot{T}^{ij}  = -2 \epsilon^{uv(i} \epsilon^{mj)p}
 \prec D_u  D_m B_{0p}, B_{0v}\succ
+ \epsilon^{mn(i}\prec [A_0, B_{0n}], D_m \Pi^{j)}  \succ+\nonumber \\
&&+\sigma^2 \Lambda \epsilon^{mn(i} <\Pi^{j)},[B_{0m},B_{0n}]>
+ \epsilon^{mn(i} \prec D_m \Pi^{j)}, \mu_{n}\succ +
 \sigma^2 \varphi^{0kl(i} <\Pi^{j)},[B_{0k},B_{0l}]> \nonumber \\
&&+ \frac{1}{4}\sigma^2 \varphi^{mnrs}\epsilon_{mnk}
\epsilon_{rsl}<\Pi^{(k} \epsilon^{l)v(i}, [\Pi^{j)},
B_{0v}]>\approx 0 \;.
\end{eqnarray}
As a result, conservation of the constraints $T^{ij}$ is a system
of  equations which will eventually fix some of the yet unfixed
Lagrange multipliers $\varphi^{klmn}.$ This point depends
crucially on invertibility properties of the following tensor:
\begin{eqnarray}\label{Dewittsupermetric}
G^{uvij}=\sigma^2 <\Pi^{(v} \epsilon^{u)r(i}, [\Pi^{j)}, B_{0r}]> \;.
\end{eqnarray}

We now show that $G^{uvij}$ can be rewritten in terms of the
Urbantke metrics.  It follows that its properties depend on the
sector under consideration.
\begin{lemma}
The tensor $G^{uvij}$ satisfies the identity:
\begin{equation}
G^{uvij}\approx
\frac{\sigma^2}{4}{\cal V}
(\epsilon^{ki(u}\epsilon^{v)lj}+
\epsilon^{kj(u}\epsilon^{v)li})(g_{kl}^{(+)}+g_{kl}^{(-)}).
\end{equation}

This implies that:

-in the topological sector (I) the matrix $G$ identically
vanishes.

-in the non topological sector  (II) the matrix $G$ is invertible
and its determinant reads:
\begin{equation}
\det(G)\approx -\frac{1}{4} {\cal V}^6 (det(g_{kl}))^2
\end{equation}
\end{lemma}

Proof:  Using the formula defining the Urbantke metrics, and the
formula expressing $B_{ij}$ in term of $\Pi$, we get:
\begin{equation}
g_{ij}^{(\epsilon)}\approx\frac{1}{\cal V}\epsilon_{abc}
\epsilon_{m(ir} B_{0j)}^{(\varepsilon)a}\Pi_{(\varepsilon)b}^{m}
\Pi_{(\varepsilon)c}^{r}.
\end{equation}
Using this result we can now compute the right handside of (\ref{Dewittsupermetric}), which is equal to $G^{uvij}.$

We have previously seen that the topological phase (I) is
characterized by the condition
$g_{\mu\nu}^{(+)}+g_{\mu\nu}^{(-)}=0.$ This implies that $G$ is
zero in this sector. In the non-topological sector we have $
G^{uvij}\approx \frac{\sigma^2}{2}{\cal V}
(\epsilon^{ki(u}\epsilon^{v)lj}+
\epsilon^{kj(u}\epsilon^{v)li})g_{kl}.$ From this expression one
easily computes the determinant $det(G)$:
\begin{equation}
det(G)\approx c {\cal V}^6 (det(g_{kl}))^2
\end{equation}
where the constant $c$ is evaluated to be $c=-\frac{1}{4}$ (go to
a frame where $g_{kl} = \delta_{kl}$). $\Box$ 

In the sequel, we
will denote by $V_s$ the spatial volume defined by $V_s^2=\vert
det(g_{kl})\vert$, so we have $det(G)\approx c {\cal V}^6 V_s^4$.

\medskip

Because the matrix $G$ is invertible in the non topological
sector, the equations $\dot{T}{}^{ij}=0$ fix the value of the 6
remaining Lagrange multipliers $\varphi^{klmn}$ and do not impose
any new constraints. By contrast, in the topological sector, these
equations do not fix these Lagrange multipliers anymore but give
tertiary constraints. The Plebanski theory turns out to define a
non-regular Hamiltonian system, i.e. its constraint analysis is
not uniform on the non-reduced phase space. In the topological
sector, we could continue the Dirac algorithm and we would observe
that the set of first class constraints is larger than in the non
topological sector. The gauge symmetries generated by this set of
first class constraints would absorb all the  local dynamical
degrees of freedom. This justifies the term ``topological'' for
this sector. In the sequel, we will exclusively concentrate on the
non topological sector.

It now remains to understand the conservation in time of the
constraints $C_0, C_i$. By anticipating the discussion concerning
separation of the constraints between first class and second
class, we will modify the constraints $C_0, C_i$ by adding to them
linear combination of constraints in such a way that the resulting
constraints, denoted  $C_0', C_i',$  are first class.

This is possible: let us define
\begin{eqnarray}
C_i'&=&C_i+<P^j,D_i B_{0j}>-<A_i,\Upsilon>-\partial_j<P^j,B_{0i}>\\
&=&<\partial_i A_j, \Pi^j>-\partial_j<A_i, \Pi^j>+
<P^j,\partial_i B_{0j}>-\partial_j <P^j, B_{0i}>.\nonumber
\end{eqnarray}
We have verified by direct computation  that $C_i'$ are indeed
first class constraints (the only non trivial computations are the
weak Poisson commuting property with respect to $T^{ij}$ and $C_0$
). Consequently, they are conserved under time evolution. As to
the constraint $C_0'$, we define it as being the total Hamiltonian
with the values of the Lagrange multipliers  fixed by the Dirac
analysis: this is precisely solving the condition that $C_0'$ is
first class with $C_0'-C_0$ a linear combination of constraints.


\subsection*{III.2 First and Second Class constraints}

\subsubsection*{III.2.1. Separation of constraints}
Once all the constraints have been determined, one must split them
into first class and second class constraints. To achieve this (in
the non topological sector), we reorganize the set of all the
constraints as follows:
\begin{eqnarray}
&&\kappa_0=<P^i,B_{0i}>\approx 0 \label{C1}\\
&&\kappa_i=\epsilon_{ijk}<P^j,\Pi^k> \approx 0\; ,\label{C2} \\
&&\Phi(B,P)^i_j=<P^{i},B_{0j}> - \frac{1}{3}
\delta^i_j <P^{k},B_{0k}>\approx 0 \label{C3}\\
&& \Phi(P,\Pi)^{ij}=<P^{(i},\Pi^{j)}>\approx 0 \label{C4}\\
&&\Upsilon=D_p\Pi^p-[P^{p},B_{0p}]\approx 0\label{C5}\\
&& \Phi(B,B)_{ij}=\prec B_{0i}, B_{0j} \succ\approx 0\label{C6}\\
&&\Phi(B,\Pi)^i_j=\prec B_{0j},\Pi^i \succ -
\frac{1}{3}\; \delta^i_j \; {\cal V} \approx 0\label{C7}\\
&&\Phi(\Pi,\Pi)^{ij}=\prec \Pi^i , \Pi^j \succ \approx 0\label{C8}\\
&&T^{ij}=\epsilon^{mn(i} \prec D_m \Pi^{j)},  B_{0n} \succ \approx 0\label{C11}\\
&& C'_i= <F_{ij},\Pi^j> - <P^j,D_i B_{0j}> + <A_i, \Upsilon>\approx 0\label{C10}\\
&&C'_0\approx 0\label{C9}
\end{eqnarray}
By a direct computation it is easy to check that
$\kappa_0,\kappa_i,\Upsilon^{(\varepsilon)a}, C'_0, C'_i$ are
first class constraints. The remaining constraints turn out to be
second class. To show that this is indeed the case, we compute
their Dirac matrix (i.e., the matrix of the Poisson brackets of
the second class constraints)and prove its invertibility.

It appears that all the Poisson brackets of the remaining
constraints are weakly ultralocal except $\{T^{ij}(x),
T^{kl}(y)\}$ which contains non ultralocal term of $\delta'(x,y)$
type. This can be taken care of (leading to simpler brackets) by
redefining $T^{ij}$ as follows:
$${\tilde T}^{ij}(x)=T^{ij}(x)+\int dy X^{(ij)}_{(kl)}(x,y)\Psi(\Pi,\Pi)^{kl}(y) $$
where $X^{(ij)}_{(kl)}(x,y)=-\frac{1}{2}\{T^{ij}(x),
T^{k'l'}(y)\}G^{-1}_{k'l'kl}(y)$, so that all the Poisson brackets
are now weakly ultralocal (i.e., contain only $\delta (x,y)$ and
none of its derivatives). Thus, if we denote by $\chi_\alpha$ the
set of redefined second class contraints, we now have
$$\{\chi_\alpha(x),
\chi_\beta(y)\}=\Delta_{\alpha\beta}(x)\delta(x,y).$$  We shall
exhibit the coefficients $\Delta_{\alpha\beta}$.

The Dirac matrix has the following structure:

\medskip
\begin{tabular}{|c||c|c|c|c|c|c|}
\hline
$\Delta_{\alpha'\alpha}$ & $\Phi(B,B)_{mn}$& $\Phi(\Pi,P)^{rs}$ &$\Phi(B,\Pi)^u_v $&
$\Phi(B,P)^j_i$& $\Phi(\Pi,\Pi)^{pq}$&$ \tilde{T}^{kl}$\\
\hline\hline
$\Phi(B,B)_{m'n'}$& 0&$ A_{m'n'}^{rs}$&0& 0&0 &0\\\hline
$\Phi(\Pi,P)^{r's'}$& $-A_{mn}^{r's'}$&0&0& $K^{jr's'}_i$&0&$M^{r's'kl}$\\\hline
$\Phi(B,\Pi)^{u'}_{v'}$& 0& 0 & 0& $B^{u'j}_{v'i}
$&0&  $N^{u'kl}_{v'}$\\\hline
$\Phi(B,P)^{j'}_{i'}$&0& $-K^{j'rs}_{i'}$&  $-B^{j'u}_{i'v}$&0&0&
$Q^{j'kl}_{i'}$\\\hline
$\Phi(\Pi,\Pi)^{p'q'}$&0& 0& 0&0&0&
$G^{p'q'kl}$\\\hline
$\tilde{T}^{k'l'}$&0& -$M^{rsk'l'}$ &$-N^{uk'l'}_{v}$&$ -Q^{jk'l'}_{i}$&$-G^{pqk'l'}$&
$0$\\\hline
\end{tabular}
\vspace{0.5cm}

\noindent where the tensor $A,B,K,M,N,Q$ can be easily written
down. In order to compute the determinant, only the values of $A$
and $B$ are necessary (besides the already known $G$) and we shall
thus give explicitly here only those. We have:
\begin{eqnarray}
&&A^{rs}_{m'n'}(x)=\frac{2}{3}{\cal V}\delta^{(r}_{(m'} \delta^{s)}_{n')}\\
&&B^{u'j}_{v'i}(x)=\frac{1}{3}{\cal V}
(\delta^{j}_{v'}\delta^{u'}_{i}-\frac{1}{3}
\delta^{j}_{i}\delta^{u'}_{v'} )
\end{eqnarray}

\begin{proposition}The determinant of the Dirac matrix is:
\begin{eqnarray}\label{determinantofDelta}
D_{2}=\det \Delta  \approx \det(A)^2 \det(B)^2 \det(G)^2 =2^8 3^{-28} {\cal V}^{40}  {V_s}^8.
\end{eqnarray}
\end{proposition}
Proof: A developpement of the determinant of the Dirac Matrix
along lines and columns give immediately the first equality. The
second equality is obtained by pluging the explicit expression of
$\det(A), \det(B), \det(G)$ which are given by:
\begin{equation}
\det(A)=(\frac{2}{3})^6 {\cal V}^6, \det(B)=(\frac{1}{3})^8 {\cal
V}^8.
\end{equation}
$\Box$

Because this determinant does not vanish, the constraints
$\chi_\alpha$ are indeed all second class.

\subsubsection*{III.2.2.  Dirac brackets}

Due to the simple structure of the Dirac matrix, the expressions
of the cofactors are easy to obtain explicitly and appear to be
polynomials in the dynamical variables and their derivatives. This
implies that we can compute exactly the inverse of the Dirac
matrix and we will denote by $\{, \}_D$ the Dirac bracket. We will
not give the explicit expression of the Dirac bracket of the
dynamical variables in the present work because no interesting
structure seems to show up.  In particular, we have seen no
immediate hint for a quantum group structure.

\subsubsection*{III.2.3.  Counting degrees of freedom}

The counting of the degrees of freedom is direct: there are 72
dynamical variables ($A_i,\Pi^i,B_{0i},P^i$),  40 second class
constraints and 14 first class constraints. Therefore, the reduced
phase space is 4 dimensional, which is correctly the number of
degrees of freedom of general relativity.

The $14$ first class constraints $\kappa_0,\kappa_i, C'_0, C'_i,
\Upsilon^{(\varepsilon)a}$ are related to the gauge symmetries of
the theory.  As was pointed out above, the constraint functions
$\Upsilon^{(\varepsilon)a}$ generate infinitesimal gauge
transformations of type ${\cal G}$. As to the constraints
$\kappa_0,\kappa_i,C'_0, C'_i$, they are related to the spacetime
diffeomorphisms.  The primary constraints $\kappa_0=0$, $\kappa_i
=0$ appear because we treat the components $B_{0i}$, which contain
the lapse $N$ and the shift $N^i$, as dynamical variables (so, the
constraints $\kappa_0=0$, $\kappa_i =0$ are equivalent to the
constraints $\pi_N = 0$, $\pi_{N^i} = 0$).  The secondary
constraints $C'_0= 0$,  $C'_i = 0$ are equivalent to the
Hamiltonian and momentum constraints.

Note that as a consequence of the constraints, the Hamiltonian $H$
weakly vanishes.  This was somehow to be expected, since the
theory is diffeomorphism-invariant.

\section*{IV. Path Integral Measure}
This section aims at  defining the path integral of the Plebanski
theory restricted to the gravitational sector. Many difficulties
have to be overcome:

1. to formulate a formal path integral taking into account the
second class constraints

2. to apply a BRST procedure to gauge fix the first class
constraints

3. to regularize and/or to discretize these path integrals

4. to ``renormalize'' these amplitudes

\noindent Our (modest) contribution will cover only the first
point. We will define the path integral measure in term  of the
Liouville measure on the phase space taking into account the
second class constraints. We will then apply the general method of
\cite{Henneaux2} to rewrite this path integral measure as a path
integral on the original configuration variables of the Lagrangian
formalism.

We first give a brief introduction to the path integral method for
pure second-class constrained systems
\cite{Senja,Faddeev,Henneaux1}, which is rather direct. We start
with a hamiltonian  system with $2n$ coordinates $(x^I)_I$ on a
symplectic phase space $({\cal P},\omega)$ and a set of $2m$
regular and irreducible second class constraints
$(\chi_\alpha)_\alpha$ . The reduced  phase space $({\cal
C},\omega_D)$ is the submanifold  ${\cal C}=\{ p \in {\cal P}
\vert \chi_\alpha(p) =0,\forall\alpha=1,\cdots,2m \}$ endowed with
the Dirac symplectic structure $\omega_D$.

The path integral is formally  defined as a sum over all paths
satisfying the constraints and the integral measure is given by
the Liouville measure associated to $\omega_D$. To be more
precise, let us introduce a complete set $y^i$ of $2(n-m)$
independent coordinates on $\cal C$ and we obtain the following
definition of the path integral:
\begin{eqnarray}
{\cal Z} \; \equiv \; \int[{\cal D}y^i(t)]
 \prod_{t\in [t_1,t_2]}\sqrt{\det\omega_D(y(t))}
 \exp (iS[y(t)]) \;.
\end{eqnarray}
In the previous formula, $S[y(t)]$ is the action of the path
$y(t)$ and $\det \omega_D$ is the determinant of the symplectic
structure.

For our purpose, it will be useful to rewrite the path integral in
terms of the original  variables $(x^I)_I$. To that end, we
perform a coordinate transformation of the non-reduced phase space
$\cal P$ from the coordinates $(x^I)_I$ to the coordinates
$(y^i,\chi_\alpha)_{i,\alpha}$. To go further, we use the fact
that it is always possible to find coordinates  $(y^i)_i$ of the
submanifold  $\cal C$ such that $\{y^i,\chi_\alpha\}= 0$. The
local symplectic structure of $\cal P$ is therefore given by the
tensor product $\omega = \omega_D \oplus \omega_\chi$ where
$(\omega_\chi^{\alpha \beta})_{\alpha,\beta}$ is the inverse of
the Dirac matrix $\Delta_{\alpha\beta} = \{\chi_\alpha,
\chi_\beta\}$ and defines the two form $
\omega_\chi=(\omega_\chi^{\alpha \beta}) d\chi_\alpha\wedge
d\chi_\beta.$

As a consequence, the equation relating the volume elements in
each coordinates system is obtained using the Liouville measure
conservation under coordinates transformations:
\begin{eqnarray}
\sqrt{\det \omega} \; \prod_{I}d x^I \; = \; \frac{\sqrt{\det \omega_D}}
{\sqrt{\det \Delta}} \;  \prod_{\alpha} d\chi^{\alpha} \;  \prod_{i} dy^i \;.
\end{eqnarray}
{}From this result, one obtains  the following expression of the
path integral in term of  the original  variables $x^I:$
\begin{eqnarray}
{\cal Z} & = & \int [{\cal D} x(t)] \; \prod_{t}( \sqrt{\det
\omega(x(t))} \; \sqrt{\det \Delta(x(t))})\nonumber \\
 && \hspace{3cm} \times \prod_{\alpha=1}^{2m}
 \delta(\chi_\alpha(x(t)))
 \; \exp(i S[x(t)])\, .
\end{eqnarray}
Thus, we see that the key point in writing the path integral is
the computation of the Dirac matrix and the evaluation of its
determinant.  This task was carried out above for the Plebanski
theory in subsection III.2.1 and the determinant was denoted
$D_2$.

In the case at hand, there are also first class constraints. These
can be gauge fixed by using $14$ gauge fixing functions
$\xi_{\alpha}$ chosen such that $D_{1}= (\det \{\psi_\alpha,
\xi_{\beta}\}_D)^2$ does not vanish on the phase space (we
collectively denote the first class constraints by
$\psi_{\alpha}$).  Once the gauge is fixed, there are only second
class constraints left and one can apply the above formula.  This
yields (for the Plebanski theory restricted to the gravitational
sector)
\begin{eqnarray}
{\cal Z}_{Pl}&=&\int [{\cal D}A_i][{\cal D}\Pi^i][{\cal
D}B_{0i}][{\cal D} P^i] \nonumber \\&& \hspace{1cm}
\exp\left(i\int dt \int_{\Sigma} \!\! d^3x \;
\left(<\Pi^i,\partial_0 A_i> + <P^i,\partial_0
B_{0i}>\right)\right) \nonumber\\&& \hspace{1cm} \prod_{x\in
M}\sqrt{D_{1}}\delta(\psi_\alpha) \delta(\xi_\alpha) \prod_{x\in
M}\sqrt{D_{2}}\delta(\Phi(B,P)) \nonumber \\ && \hspace{1cm}
\times \; \delta(\Phi(\Pi,P)) \delta(\Phi(B,B))
\delta(\Phi(B,\Pi)) \delta(\Phi(\Pi,\Pi))\delta(\tilde{T})
\label{key1}
\end{eqnarray}
Recall that the Hamiltonian vanishes when the constraints are
taken into account.  The gauge fixing conditions $\xi_\alpha$
depend on the canonical variables and can also involve explicitly
the time (``time dependent canonical gauge conditions").  For the
subsequent discussion, it will be simpler to take gauge conditions
that do not involve the potential $A_i$.  This is equivalent to
taking gauge conditions that involve the tetrads only (and their
derivatives), but not the connection.  The boundary conditions to
be imposed on the paths that are summed over in the path integral
should be dealt with along the lines of \cite{HTV}.

We may re-express (\ref{key1}) as
\begin{eqnarray}
{\cal Z}_{Pl}&=&\int [{\cal D}A_i][{\cal D}\Pi^i][{\cal
D}B_{0i}][{\cal D} P^i] \nonumber \\&& \hspace{1cm}
\exp\left(i\int dt \int_{\Sigma} \!\! d^3x \;
\left(<\Pi^i,\partial_0 A_i> + <P^i,\partial_0 B_{0i}> -
\tilde{H}\right)\right) \nonumber\\&& \hspace{1cm} \prod_{x\in
M}\sqrt{D_{1}}\delta(\psi_\alpha) \delta(\xi_\alpha) \prod_{x\in
M}\sqrt{D_{2}}\delta(\Phi(B,P)) \nonumber \\ && \hspace{1cm}
\times \; \delta(\Phi(\Pi,P)) \delta(\Phi(B,B))
\delta(\Phi(B,\Pi)) \delta(\Phi(\Pi,\Pi))\delta(\tilde{T})
\label{key2}
\end{eqnarray}
since $\tilde{H}$, defined as \begin{equation} - \tilde{H} =
\epsilon^{mnp}<B_{0p},F_{mn}>+ \Lambda \prec
B_{0p},\Pi^{p}\succ,\end{equation} vanishes on the constraint
surface.

Our aim is to rewrite this expression as a path integral over the
original covariant configuration variables with the original
Plebanski action. To proceed, we first integrate over the momenta
$P^i.$ This can be done by using the following change of
variables:
\begin{equation}
[{\cal D}P^i]=J[{\cal D}\Phi(B,P)][{\cal D}\Phi(\Pi,P)] [{\cal
D}\kappa_0][{\cal D}\kappa_i]
\end{equation}
where $J$ is the Jacobian of the linear transformation $P^i\mapsto
(\Phi(B,P),\Phi(\Pi,P),\kappa_0,  \kappa_i)$, easily shown to be
equal to $J=(\prec B_{0k},\Pi^k \succ)^{-9}={\cal V}^{-9}.$ As a
result we obtain:
\begin{eqnarray}
{\cal Z}_{Pl}&=&\int [{\cal D}A_i][{\cal D}\Pi^i][{\cal
D}B_{0i}]\nonumber \\&& \hspace{1.5cm}\exp\left(i \int dt
\int_{\Sigma} \!\! d^3x \;\left(
<\Pi^i,\partial_0 A_i> - \tilde{H}  \right)\right)\nonumber\\
&& \hspace{1.5cm}\prod_{x\in
M}\sqrt{D_{1(P_i=0)}}\delta(C'_0)\delta(C'_i)\delta(\Upsilon)
\delta(\xi_\alpha) \nonumber \\ && \hspace{1.5cm} \times \,
\sqrt{D_{2}}{\cal V}^{-9} \delta(\Phi(B,B)) \delta(\Phi(B,\Pi))
\delta(\Phi(\Pi,\Pi))\delta(\tilde{T}). \label{key3}
\end{eqnarray}

We now apply the method of \cite{Henneaux2} to eliminate the
secondary second class constraints $T^{ij}$. We denote
$\zeta_{1}^{ij}=-\frac{1}{4}\Phi(\Pi,\Pi)^{ij}$ as well as
$\zeta_{2}^{ij}=T^{ij}.$ The constraints $\zeta_{2}^{ij}$ are the
secondary constraints associated with $\zeta_{1}^{ij}$ in the
sense that we have $\{\zeta_{1}^{ij},\tilde{H}\}=\zeta_{2}^{ij}$.
Introducing multipliers $\lambda_{ij}$ and $\rho_{ij}$ to enforce
the constraints, we can rewrite (\ref{key3}) as:
\begin{eqnarray*}
&&{\cal Z}_{Pl}=\int [{\cal D}A_i][{\cal D}\Pi^i][{\cal D}B_{0i}]
[{\cal D}\lambda_{ij}][{\cal D}\rho_{ij}]\\
&&\exp\left(i \int dt \int_{\Sigma} \!\! d^3x \;\left(
 <\Pi^i,\partial_0 A_i>- \tilde{H} -
\lambda_{ij}\zeta_{1}^{ij}-\rho_{ij}\zeta_{2}^{ij}\right)\right)\\
&&
 \prod_{x\in M}\sqrt{D_{1(P_i=0)}}\delta(C'_0)\delta(C'_i)\delta(\Upsilon)
\delta(\xi_\alpha)
\sqrt{D_{2}}{\cal V}^{-9}
 \delta(\Phi(B,B)) \delta(\Phi(B,\Pi)).
\end{eqnarray*}
We consider a canonical transformation generated by the function
$F=-\rho_{ij}\zeta_1^{ij}.$ Under this transformation the kinetic
term $<\Pi^i,\partial_0 A_i>$ is left invariant\footnote{The
canonical transformation depends explicitly on the parameter
$\rho_{ij}$, so the kinetic term transforms with the explicit time
derivative of the generator, $\delta <\Pi^i,\partial_0 A_i> \sim
\dot{\rho}_{ij}\zeta_1^{ij}$.  Since the constraint $\zeta_1^{ij}$
remains invariant (and enforced through the integration over
$\lambda_{ij}$), the kinetic term is indeed invariant on the
constraint surface.  Alternatively, its variation in the
non-reduced phase space can be absorbed through a redefinition of
$\lambda_{ij}$. }, as well as the path integral measure and also
$\lambda_{ij}\zeta_{1}^{ij},$ whereas $\tilde{H}$ is turned into
$$\tilde{H}'=\exp(\{F,.\})(\tilde{H})=
\tilde{H}-\rho_{ij}\zeta_2^{ij}-2\rho_{ij}\rho_{kl}G^{ijkl}$$ and
$\zeta_2^{ij}$ is turned into
$${\zeta'}_2^{ij}=\zeta_2^{ij}+4\rho_{kl}G^{ijkl}$$
(to obtain the latter results, we have used $\{\zeta_1^{ij},
\zeta_2^{kl}\}=-4G^{ijkl}$ and $\{G^{ijkl}, F\}=0$). Finally,
since only $A_i$ transforms under the canonical transformation
generated by $F$, the product of the $\delta$-distributions in the
path integral is invariant (we assume that the gauge conditions do
not involve $A_i$). As a result, we obtain
\begin{eqnarray*}
&&{\cal Z}_{Pl}=\int [{\cal D}A_i][{\cal D}\Pi^i][{\cal D}B_{0i}]
[{\cal D}\rho_{ij}]\\
&&\exp\left(i \int dt \int_{\Sigma} \!\! d^3x \;\left(
 <\Pi^i,\partial_0 A_i>- \tilde{H}-2\rho_{ij}
\rho_{kl}G^{ijkl}\right)\right)\\
&&\prod_{x\in
M}\sqrt{D_{1(P_i=0)}}\delta(C'_0)\delta(C'_i)\delta(\Upsilon)
\delta(\xi_\alpha) \sqrt{D_{2}}{\cal V}^{-9}
 \delta(\Phi(B,B)) \delta(\Phi(B,\Pi))\delta(\Phi(\Pi,\Pi)).
\end{eqnarray*}
The integration over the variables $\rho_{ij}$ is gaussian and
gives
\begin{eqnarray*}
&&{\cal Z}_{Pl}=\int [{\cal D}A_i][{\cal D}\Pi^i][{\cal D}B_{0i}]
\exp\left(i \int dt \int_{\Sigma} \!\! d^3x \;\left(
 <\Pi^i,\partial_0 A_i>- \tilde{H}\right)\right)\\
&&\prod_{x\in
M}\sqrt{D_{1(P_i=0)}}\delta(C'_0)\delta(C'_i)\delta(\Upsilon)
\delta(\xi_\alpha) \frac{\sqrt{D_{2}}{\cal V}^{-9}}{\sqrt{det(G)}}
 \delta(\Phi(B,B)) \delta(\Phi(B,\Pi))\delta(\Phi(\Pi,\Pi)).
\end{eqnarray*}
We have thus succeeded in eliminating the constraints $T_{ij}$.

To make contact with the Plebanski action, we now reintroduce the
original configuration variables. Because
$\Pi^i=\epsilon^{ijk}B_{jk}$, the Jacobian of the transformation
from $\Pi^i$ to $B_{jk}$ is a constant, therefore $[{\cal
D}\Pi^k]=[{\cal D}B_{ij}]$. We can furthermore eliminate the Dirac
distributions $\delta(\Upsilon), \delta(\Phi(B,B)),
\delta(\Phi(B,\Pi)),\delta(\Phi(\Pi,\Pi))$ by restoring the
integration over $A_0$ and $\varphi^{\mu\nu\rho\sigma}$. Taking
into account the known values of $\sqrt{D_{2}}$ and
$\sqrt{det(G)}$, we obtain the final expression:
\begin{eqnarray}
&&{\cal Z}_{Pl}=\int [{\cal D}A_{\mu}][{\cal D}B_{\mu\nu}] [{\cal
D}\varphi^{\mu\nu\rho\sigma}]{\cal V}^8 V_s^2 \exp{(i S_{Pl}[A, B,
\varphi])}\nonumber \\ && \hspace{2cm} \prod_{x\in
M}\sqrt{D_{1(P_i=0)}}\delta(C'_0)\delta(C'_i) \delta(\xi_\alpha) .
\end{eqnarray}

This is the formal expression of the gauge fixed measure of
Plebanski theory in the non topological sector. One can be puzzled
by the fact that the measure has an expression which is non
covariant because it contains $V_s^2.$ This phenomenon is typical
of derivative couplings and is completely analogous to the non
covariant measure computed 40 years ago by H.Leutwyler
\cite{Leutwyler} which reads:
\begin{equation}
{\cal D}g= \prod_x {\cal V}^{-5}V_s^2 \prod_{\mu\leq \nu}dg_{\mu\nu}
 \end{equation}
A derivation of this measure from the Liouville measure was done
by  E.S.Fradkin and G.A.Vilkovisky  \cite{FV,Fradkin} as well as a
careful analysis of the covariance of the correlation functions.
See also \cite{KS} for the different comparisons of functional
measure for quantum gravity.

\section*{V. Conclusion}
In the present work we have derived the hamiltonian analysis of
the Plebanski theory.  A notable feature of this theory is that it
defines a non-regular theory, in the sense that the matrix of the
Poisson brackets of the constraints does not have constant rank on
the constraint surface.  Non-regular theories used to be thought
of as being somewhat pathological and of no interest to physical
systems.  In the last years, however, models have been constructed
that lead to non-regular constrained Hamiltonian  systems. Another
example, besides Plebanski theory, is given by the
higher-dimensional pure Chern-Simons theories \cite{BGH1,BGH2}. A
recent analysis of non-regular systems may be found in
\cite{MZ}.

We have explicitly carried out the derivation of the Hamiltonian
formalism in the non topological sector. In particular, we have
obtained all the constraints and been able to disentangle them
into first class and second class. We have also computed exactly
the local measure, which has enabled us to give a formal path
integral of Plebanski theory (in the non topological sector) in
term of the original variables. This constitutes a first step in
the direction of the understanding of Plebanski theory.

Our work should be further extended in various directions. First,
the above path integral was written in a canonical gauge.  It
would be useful to derive it in propagating gauges by BRST
methods. 

 Second, it would also be of interest to try to
discretize the Plebanski theory using spin foam models.
A central issue is the computation of the weights of the faces and edges
 which are still a matter of debate \cite{BCHT,FL,BP}.
The computation of the functional measure that we provide will hopefully give new insights in this intricate problem.

Finally, one of our motivations was to try to uncover a quantum
group structure underlying the Plebanski theory. In euclidean $BF$ theory with cosmological constant it is well known that quantum groups are hidden in this topological field theory \cite{CY}. Up to now there is no clear picture of appearance of quantum groups (more precisely of Poisson Lie groups or Poisson Lie groupoids) as symmetries of Poisson brackets of the fields in this theory.  
 Unfortunately, the form
of the Dirac brackets that we can compute in Plebanski theory does not appear to be connected with such 
structures in any obvious way.
 The relation, if any,  between quantum groups and Plebanski theory still remains to us a  mystery as well as the 
relation between   the quantum deformation of the Barrett-Crane model \cite{NR} and  a discretization of Lorentzian Plebanski theory with cosmological constant.

\section*{Acknowledgments}

M.H. is grateful to the ``Laboratoire de physique math\'ematique
et th\'eorique" of the University of Montpellier for kind
hospitality while this work was carried out. The research of M.H.
is supported in part by the ``Actions de Recherche Concert{\'e}es"
of the ``Direction de la Recherche Scientifique - Communaut{\'e}
Fran{\c c}aise de Belgique", by a ``P\^ole d'Attraction
Interuniversitaire" (Belgium) and by IISN-Belgium (convention
4.4505.86). Support from the European Commission RTN programme
HPRN-CT-00131, in which M.H. is associated to K. U. Leuven, is
also acknowledged.

The research of K.N. is supported by NSF grants PHY-0090091 and the Eberly Research Funds of Penn State University.

\newpage

\section*{Appendix}
\subsection*{Algebra $so(\eta)$: definition, notations and properties}
$g=so(\eta)$ is the real Lie algebra of the isometry group
$G=SO(\eta)$ of the quadratic form
$\eta=\text{diag}(\sigma^2,1,1,1)$ where  $\sigma\in \{1,i\}$. We
will use capital latin letters for internal vector indices
$I,J,\cdots \in \{0,1,2,3\}$ and we will define the antisymmetic
tensor $\epsilon^{IJKL}$ such that $\epsilon^{0123}=1$; the
indices are lowered and raised with the metric $\eta$.

One of the basis of $so(\eta)$, composed of {\it rotation}
generators $(J_a)_{a=1,2,3}$ and  {\it boost} generators
$(K_a)_{a=1,2,3}$ has the following  commutation relations:
\begin{eqnarray}\label{algebra}
[J_a,J_b]=\epsilon_{ab}{}^c J_c \;\; , \;\; [K_a,K_b]
= \sigma^2 \epsilon_{ab}{}^c J_c \;\; , \;\; [K_a,J_b]=\epsilon_{ab}{}^c K_c \;.
\end{eqnarray}
We have used the notation $\epsilon_{abc}=\epsilon^{0}{}_{abc}.$

The real Lie algebra $so(\eta)$ is also a real form of
$so(4,\mathbb{C})=so(3,\mathbb{C})\oplus {so(3,\mathbb{C})}$.

It is convenient to introduce the basis $(T_{_{}^{{(\pm)
a}}})_{a=1,2,3}$ defined from the previous one by $T_{_{}^{(\pm)
a}}=\frac{1}{2}(J_a \pm \sigma K_a)$ whose generators realize two
commuting copies of $so(3)$, i.e.:
\begin{eqnarray}\label{basisTpm}
[T_{{}^{(\pm) a}}, T_{{}^{(\pm) b}}] = \epsilon_{ab}{}^c
\; T_{{}^{(\pm) c}} \;\;\; , \;\;\;  [T_{{}^{(+)a}}, T_{{}^{(-)b}}] = 0 \;.
\end{eqnarray}
We denote by $p^+$ and $p^-$ the projectors associated to this direct sum.
To any element  $\xi$ of $so(4,\mathbb{C})$, we denote
$\xi^{\pm}=p^{\pm}(\xi)= \xi^{^{_{(\pm)a}}} T_{{}^{(\pm)a}}.$

The star structure on $so(4,\mathbb{C})$ selecting the real form  reads:
\begin{equation}
\star(T_{{}^{(\pm) a}})=-T_{{}^{(\sigma^2\pm) a}}.
\end{equation}

As usual it is customary to define the Hodge  linear map
$\iota=\frac{1}{\sigma}(p^+ -p^-)$ which satisfies $\iota:
so(\eta)\rightarrow so(\eta)$ and $\iota^2=\sigma^2 id$, as well
as:
\begin{eqnarray}
\forall \; \xi, \chi \; \in \; so(\eta)
\;\;,\;\;\;\; [\iota(\xi),\chi] \; = \; [\xi, \iota(\chi)]\;.
\end{eqnarray}

The space of  real symmetric invariant bilinear forms on the Lie
algebra $so(\eta)$ is a two-dimensional vector space; we choose
two independent nondegenerate bilinear  forms
$<\!\!\cdot,\cdot\!\!> $ and $\prec \!\! \cdot,\cdot \!\! \succ$
defined by:
\begin{eqnarray}\label{Killingforms}
 <T_{{}^{(\varepsilon) a}}, T_{{}^{(\varepsilon') b}}\!>
 = \delta_{ab} \; \delta_{\varepsilon \varepsilon'} \;\;\;\; ,
 \;\;\;\; \prec T_{{}^{(\varepsilon) a}},
 T_{{}^{(\varepsilon') b}}\succ = \sigma \varepsilon
 \; \delta_{ab} \; \delta_{\varepsilon \varepsilon'}\;.
\end{eqnarray}
From the definition  of $\iota$, we have:
\begin{eqnarray}\label{swichproperty}
\forall \; \xi, \chi \; \in \; so(\eta) \;\;, \;\;\;\;
\prec \iota(\xi), \chi \succ \; = \; \prec \xi , \iota(\chi)
\succ \; = \;<\xi, \chi> \;.
\end{eqnarray}

The vector representation $\Pi$ of the Lie algebra $so(\eta)$ is
given by:
\begin{eqnarray}\label{vectorialrepresentation}
\Pi(T_{{}^{(\pm) a}})^{IJ}=\; T_{{}^{(\pm) a}}^{IJ} \;
= \; \frac{1}{2}
\left( -\epsilon^{0aIJ} \pm \sigma (\eta^{aI} \eta^{0J}
- \eta^{0I}\eta^{aJ})\right) \;.
\end{eqnarray}

To prove that the formulae (\ref{vectorialrepresentation}) define
a representation of the Lie algebra, it is sufficient to verify
the following useful relations:
\begin{eqnarray}
T_{{}^{(\varepsilon) a}}^{IJ} \eta_{JJ} T_{{}^{(\varepsilon') b}}^{JK}
\; = \; \delta_{\varepsilon \varepsilon'} \left( \frac{1}{2}
\epsilon_{ab}{}^c \; T_{{}^{(\varepsilon) c}}^{IK} -
\frac{1}{4} \eta_{ab} \delta^{IK} \right)\;.
\end{eqnarray}
Furthermore, the values of the quadratic Casimir tensors of
$so(\eta)$ in the vector representation can be computed using the
relation\begin{eqnarray} \sum_a T_{{}^{(\varepsilon) a}}^{IJ}
T_{{}^{(\varepsilon) a}}^{KL} \; = \; \frac{1}{4} \varepsilon
\sigma \epsilon^{IJKL} + \frac{1}{4} (\delta^{IK} \delta^{JL} -
\delta^{IL} \delta^{JK}) \;.
\end{eqnarray}
{}From these quadratic formulae, it is easy to compute the
components $\xi^{(\pm)a}$ of any element $\xi$ in the Lie algebra
in term of $\xi^{IJ}$ and one obtains:
\begin{eqnarray}
\xi^{(\pm)a} \; = \; - \eta^{ab} \xi^{IJ} \; \eta_{IK} \;
\eta_{KL} \; T_{{}^{(\pm) b}}^{KL} \;.
\end{eqnarray}

\bibliographystyle{unsrt}

\end{document}